\documentclass[aps,pre,twocolumn,notitlepage,superscriptaddress]{revtex4-1}

\usepackage{amsmath}
\usepackage{amssymb}
\usepackage{hyperref}
\usepackage{graphicx}
\usepackage[arrow, matrix, curve]{xy}
\usepackage{bbm}
\usepackage{bm}
\usepackage{yhmath}
\usepackage{color}
\usepackage{physics}
\usepackage{bbold}

\usepackage{color}

\renewcommand\vec[1]{\boldsymbol{\mathrm{#1}}}

\usepackage[usenames,dvipsnames]{xcolor}
\hypersetup{colorlinks=true, linkcolor=BrickRed, urlcolor=blue!50!black, citecolor=blue!50!black}

\usepackage[normalem]{ulem}

\makeatletter
\newcommand\hide@visible[1]{%
  \bgroup\fboxsep=.3ex\colorbox{Gray}{begin hide}%
  #1\colorbox{Gray}{end hide}\egroup%
}
\newcommand\hide@hidden[1]{%
  \bgroup\fboxsep=.3ex\colorbox{Gray}{hidden text}%
}
\newcommand\hide@invisible[1]{}
\newcommand\makevisible{\let\hide\hide@visible}
\newcommand\makehidden{\let\hide\hide@hidden}
\newcommand\makeinvisible{\let\hide\hide@invisible}
\makeatother
\makehidden

\begin{document}

\title{Learning how to find targets in the micro-world: \\The case of intermittent active Brownian particles}

\author{Michele Caraglio}
\email{Michele.Caraglio@uibk.ac.at}
\affiliation{Institut f\"ur Theoretische Physik, Universit\"at Innsbruck, Technikerstra{\ss}e 21A, A-6020, Innsbruck, Austria}

\author{Harpreet Kaur}
\affiliation{Institut f\"ur Theoretische Physik, Universit\"at Innsbruck, Technikerstra{\ss}e 21A, A-6020, Innsbruck, Austria}

\author{Lukas J. Fiderer}
\affiliation{Institut f\"ur Theoretische Physik, Universit\"at Innsbruck, Technikerstra{\ss}e 21A, A-6020, Innsbruck, Austria}

\author{Andrea L{\'o}pez-Incera}
\affiliation{Institut f\"ur Theoretische Physik, Universit\"at Innsbruck, Technikerstra{\ss}e 21A, A-6020, Innsbruck, Austria}

\author{Hans J. Briegel}
\affiliation{Institut f\"ur Theoretische Physik, Universit\"at Innsbruck, Technikerstra{\ss}e 21A, A-6020, Innsbruck, Austria}

\author{Thomas Franosch}
\affiliation{Institut f\"ur Theoretische Physik, Universit\"at Innsbruck, Technikerstra{\ss}e 21A, A-6020, Innsbruck, Austria}

\author{Gorka Mu{\~n}oz-Gil}
\affiliation{Institut f\"ur Theoretische Physik, Universit\"at Innsbruck, Technikerstra{\ss}e 21A, A-6020, Innsbruck, Austria}

\date{\today}

\begin{abstract}
Finding the best strategy to minimize the time needed to find a given target is a crucial task both in nature and in reaching decisive technological advances.
By considering learning agents able to switch their dynamics between standard and active Brownian motion, here we focus on developing effective target-search behavioral policies for microswimmers navigating a homogeneous environment and searching for targets of unknown position.
We exploit \textit{Projective Simulation}, a reinforcement learning algorithm, to acquire an efficient stochastic policy represented by the probability of switching the phase, i.e. the navigation mode, in response to the type and the duration of the current phase.
Our findings reveal that the target-search efficiency increases with the particle's self-propulsion during the active phase and that, while the optimal duration of the passive case decreases monotonically with the activity, the optimal duration of the active phase displays a non-monotonic behavior.
\end{abstract}

\maketitle

\section*{Introduction}

Target search is a universal problem occurring in various fields and at several length scales~\cite{Benichou2011}.
Examples range from animals searching for food, mate, or shelter~\cite{Charnov1976,Obrien1990,Sims2008,Kramer2015} to castaway rescue operations~\cite{Frost2001}, and to proteins binding to specific DNA sequences~\cite{Berg1981,Gorman2008}.
At the micro-scale, further paradigmatic examples include bacteria foraging nourishment~\cite{Elgeti2015,Berg2004}, phagocytes of the immune system performing chemotactic motion during infection~\cite{Devreotes1988,Deoliveira2016}, and sperm cells on their way to the egg~\cite{Eisenbach2006}.
Furthermore, artificial and biohybrid microswimmers~\cite{Smanski2016,You2018,Klumpp2019} with good target-search skills have been envisaged as revolutionary smart materials able to perform assisted fertilization~\cite{Medina-Sanchez2016}, targeted drug delivery~\cite{Patra2013,Naahidi2013}, or environmental remediation~\cite{Gao2014}.

In many relevant circumstances, the agent has no \textit{a priori} knowledge of the target location and has to develop effective stochastic strategies that allow minimizing, at least on average, the search time in an environment with randomly distributed targets.
Motivated by observational data, physical intuition, and analytical tractability, L\'evy walks~\cite{Viswanathan2011,Viswanathan1999,Viswanathan2008} and intermittent searches~\cite{Benichou2011,Benichou2005,Benichou2006,Loverdo2009} are among the statistical strategies that have received major attention in the past.
In the former, the agent undergoes straight runs at constant speed with run lengths $l$ drawn from a L\'{e}vy distribution $p(l) \propto l^{-(\alpha+1)}$, with $0<\alpha<2$ and the target is detected if the searcher transits closer than a threshold distance, which also acts as a small-lengths cut-off allowing to normalize the L\'{e}vy distribution.
The optimal value of $\alpha$ depends sensitively on model details such as the revisitability and mobility of the targets or the complexity of the environment~\cite{Viswanathan1999,Santos2004,Bartumeus2002,Volpe2017}.
Intermittent-search strategies have been proposed based on the observation that fast movements allow exploring quickly the whole environment but may, on the other hand, significantly degrade perception abilities~\cite{Obrien1990}.
In these strategies, phases of diffusive motion permitting target detection are alternated with phases of ballistic motion which allow quick relocation to different positions at the cost of not being able to detect the target.
In the simplest version of the model, the agent switches from one phase to the other with a fixed rate leading to exponentially distributed phase durations, but other distributions have also been considered~\cite{Benichou_2007,Lomholt2008}.
The mean search time of these strategies can be minimized under broad conditions.
In particular, it has been shown that there is an optimal duration of the ballistic nonreactive phase which depends only on the dimensionality of the system and is independent of the details of the slow reactive phase~\cite{Loverdo2009}.
Intermittent-search strategies remain robust also in the cases of different target distributions such as patches~\cite{Benhamou1992} and Poissonian distributions in one dimensions~\cite{Moreau2009}.

In the past decade, machine learning has emerged as a revolutionary tool helping to elucidate various aspects of active matter systems~\cite{Cichos2020}.
In particular, reinforcement learning (RL)~\cite{Sutton2018} and genetic algorithms~\cite{Mitchell1998} have proved to be powerful tools able to identify successful swimming strategies improving the navigation performances of microswimmers and their odds of reaching a target.
Promising and worthy results have been obtained in several situations including simple energy landscapes~\cite{Schneider2019}, viscous surroundings~\cite{Muinos-Landin2021,Tsang2020,Hartl2021}, complex motility fields~\cite{Monderkamp2022}, and steady or turbulent flows~\cite{Colabrese2017,Gustavsson2017,Colabrese2018,Biferale2019,Alageshan2020}.
However, previous literature has mainly focused on increasing the net flux of particles in a certain direction or on optimizing point-to-point navigation towards a target whose position is fixed and then implicitly learned during the learning process.
Thus, notwithstanding the increasing popularity of machine-learning algorithms in the active matter field~\cite{Cichos2020,Tsang2020review} and the previously mentioned seminal works, investigation of stochastic target-search problems with randomly distributed targets via machine-learning approaches remains largely unexplored, with only a couple of very recent exceptions~\cite{munozgil2023,Kaur2023}.

Mu{\~n}oz-Gil \textit{et al.}~\cite{munozgil2023} have applied RL methods to learn optimal foraging strategies outperforming the efficiency of L\'evy walks in the case of revisitable, sparsely distributed targets.
In their setup, the learning agent performs a stepwise motion with constant velocity and, at each step, decides if maintaining the current direction or turning in a new random one, this choice being based only on the length of the current straight segment.
Whenever the agent detects a target, it receives a reward and, through several trials, it optimizes its policy, learning an efficient distribution of the length of the straight segments.
The former approach, with respect to traditional analytical ones, has the advantage of not being restricted to a specific ansatz of the straight-segment length distribution.
However, it remains focused on investigating known idealized scenarios, which are not entirely apt for describing the behavior of real microswimmers.

On the other hand, Kaur \textit{et al.}~\cite{Kaur2023} rely on the active Brownian particle (ABP)~\cite{Bechinger2016}, which well describes the behavior of artificial microswimmers, and show that genetic algorithms manage to address the problem of finding targets of unknown positions for particles able to decide if and when switching their behavior between standard passive Brownian diffusion and directed ABP motion.
In particular, they use the algorithm NeuroEvolution of Augmenting Topologies~\cite{Stanley2002} to evolve an initial population of particles taking random decisions towards a population in which the majority of particles are optimized to solve the target-search problem.
However, their findings are limited by the fact that, in their setup, a given individual particle acts deterministically in the sense that it always selects the same duration for each phase from a set of predetermined durations.

In the present manuscript, we combine the two former approaches: 
While, as in Ref.~\cite{Kaur2023}, we resort again on agents able to switch their behavior between passive and active Brownian motion, we here exploit the powerful RL framework employed in Ref.~\cite{munozgil2023}, thus allowing our agents to learn a distribution of durations for each of the two phases.
This results in a stochastic strategy maximizing the foraging efficiency in a homogeneous environment, which can eventually be tested in experiments with artificial Janus particles~\cite{Howse2007,Jiang2010} where the activity is controlled by an external illuminating system~\cite{Muinos-Landin2021}.

\section*{Model}

With intermittent-search strategies in mind, we design our agent as a particle switching between two different phases $\phi$ and able to keep track of the current phase duration $\omega$.
More specifically, the particle can perform either standard Brownian diffusion ($\phi=0$) enabling target detection, or active Brownian motion ($\phi=1$), which does not allow to sense the target but, depending on its self-propulsion, may quickly relocate the particle to a different region.
In the following, the two navigation modes are also referred to in short as the Brownian Particle (BP) phase and the ABP phase, respectively.
At each time $t$, the \textit{state} of the agent $s_t$ is then characterized by the tuple $s_t=(\phi_t, \omega_t)$, with $\phi_t$ representing the current phase and $\omega_t$ its time duration since the last switching event.
As customary in the RL framework~\cite{Sutton2018}, given its current state $s_t$, the agent replies with an \textit{action} $a_t$, and gains a \textit{reward} if this action leads to a benefit for the agent.
In our case, the action corresponds to making a decision on whether to maintain the current phase or switch to the other one.
This choice follows a probabilistic rule, with $p_t$ the probability of switching phase.
We highlight the fact that, in our approach, $p_t$ is not a constant but it depends on the current state $s_t$ of the agent.
The full set of these probabilities (one for each state $s_t$) constitutes the \textit{policy} of the agent.
During the learning process, such a policy is constantly updated with the goal of maximizing the total reward, i.e. the target-search efficiency (see Methods section for more details).

Including these notions into the standard ABP model~\cite{Bechinger2016} in a homogeneous environment results in the following set of Langevin equations, discretized according to It\^{o} rule,
\begin{eqnarray}\label{eom1}
\phi_{t+\Delta t} & = & \left\lbrace 
\begin{array}{ll}
\phi_t & \mbox{with probability } 1-p_t \; , \\
1-\phi_t & \mbox{with probability } p_t \; ,
\end{array}
\right. \\
\label{eom2}
\vec{r}_{t\!+\!\Delta t} &=& \vec{r}_{t} + v \, \vec{u}_{t} \, \phi_{t} \, \Delta t  + \sqrt{2D\Delta t} \, \boldsymbol{\xi}_t \; , \\ 
\label{eom3}
\vartheta_{t\!+\!\Delta t} &=& 
\left\lbrace 
\begin{array}{ll}
\vartheta_{t} + \sqrt{2D_{\vartheta}\Delta t} \, \eta_t  & \mbox{if } \phi_{t+\Delta t} = \phi_t  \; ,\\
2\pi \mbox{ \textbf{rand}} & \mbox{otherwise} \; .
\end{array}
\right. 
\end{eqnarray}
Here $\Delta t$ is the integration time step, $\vec{r}_t = (x_t,y_t)$ is the position at time $t$, and $\vec{u}_{t} = \big(\cos\vartheta_{t},\sin\vartheta_{t}\big)$ denotes the instantaneous orientation of the self-propulsion velocity with constant modulus $v$.
$D$ and $D_{\vartheta}$ are the translational and rotational diffusion coefficients, respectively.
Finally, the components of the vector noise $\boldsymbol{\xi}_{t}=(\xi_{x,t},\xi_{y,t})$ and of the scalar noise $\eta_t$ are independent random variables, distributed according to a Gaussian with zero average and unit variance.
Note that when the phase of the particle is that of a passive Brownian particle ($\phi_t=0$), the spatial evolution is decoupled from the orientational diffusion of the self-propulsion.

Our homogeneous environment is modeled as a two-dimensional square box of size $L \times L$ with periodic boundary conditions.
A circular target of radius $R=0.05L$ is located randomly inside the box.
Every time the agent finds this target (i.e. the distance between the center of the target and the particle position is smaller than the target radius $R$), it gets a positive reward, the target is destroyed, and a new target appears at a new random location inside the box.
Due to the periodic boundary conditions, this environment is formally equivalent to an infinite domain with a lattice of targets.

\begin{figure*}[t]
\includegraphics[width=1.0\textwidth]{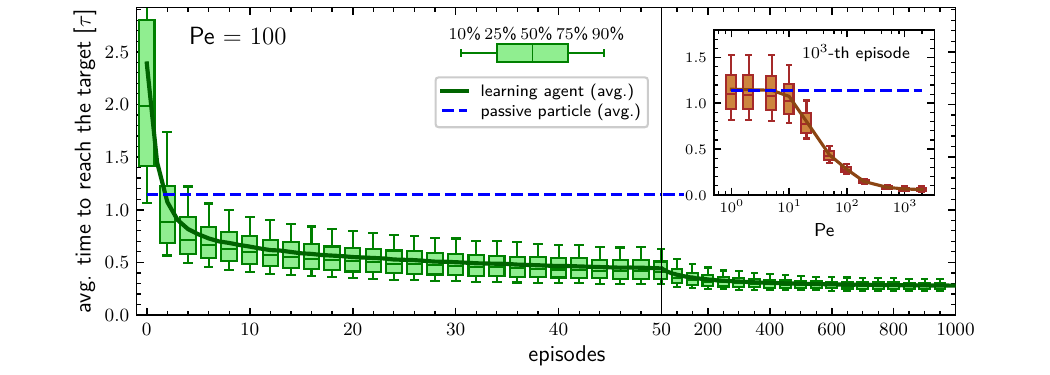}
\caption{Average time required to reach the target during an episode of duration $20\tau$ as a function of the number of episodes for $\text{Pe}=100$.
The continuous green line represents the average over $N=5\cdot10^3$ independent particles, while the box-and-whiskers symbols report respectively the 10th, 25th, 50th, 75th, and 90th percentiles.
The blue dashed line represents the average search time of a completely passive particle.
Inset: Time required to reach the target at the $10^3$-th episode as a function of the P\'eclet number for $\ell^*=1$.
The hyperparameters of the PS algorithm for each P\'eclet number are reported in the Methods section.}
\label{fig:1}
\end{figure*}

In the following, we fix the length unit as the size of the box $L$ and the time unit as the typical time $\tau := L^2/4D$ required by a passive particle to cover this distance.
The model has thus two free dimensionless parameters: The P\'{e}clet number $\text{Pe} := v\tau/L$, measuring the magnitude of the activity, and the persistence $\ell^* := v/D_{\vartheta}L$, representing the persistence of directed motion in the ABP phase.

\section*{Results}

Resorting on the RL algorithm Projective Simulation~\cite{Briegel2012} (PS) described in the Methods section, the learning performances of our agents are evaluated, for each set of free parameters, by checking how the average time to reach the target (also known as the mean first-passage time~\cite{Redner2001,Metzler2013}) evolves during subsequent episodes of duration $20\tau$.
The average is performed over $N$ independent agents, all with the same initial policy in which, independently of the current phase duration, the probabilities of switching phase are $10^{-2}$ and $10^{-3}$ when being in the passive and in the active phase respectively.
Such initial policy is purposely chosen to have particles with a rather poor initial target-search efficiency, with an average searching time at least twice that of a pure passive particle, the latter being about $1.14\tau$.

We first consider the case in which the learning particle, when in the ABP phase, has a large activity and a persistence length equal to the box size.
To do so, we set the persistence to $\ell^*=1$ and the P\'eclet number to $\text{Pe}=100$ which means that the ratio between the typical length traveled because of the self-propulsion and the typical length traveled due to diffusion is $1$ at the minimal phase duration, corresponding to the integration time step $\Delta t = 10^{-4}\tau$, and grows up to $100$ for a phase duration equal to the time unit $\tau$ (see Methods section for details).
In such a situation, the learning particle outperforms the target-search performances of a purely passive particle already after two episodes, see Figure~\ref{fig:1}.
During subsequent episodes, the average time required to find the target keeps decreasing, following a stretched exponential behavior that depends on the details of the learning algorithms (see Methods section), and after $10^3$ episodes it is about $4$ times smaller than the benchmark value corresponding to the fully passive particle.
Furthermore, also the spread of the average search times among the $N$ different agents decreases during the learning process, with the difference between the first quartile and the third quartile reducing from about $1.4\tau$ to about $0.1\tau$ along the $10^3$ episodes, see Figure~\ref{fig:1}.

\begin{figure*}[t]
\includegraphics[width=1.0\textwidth]{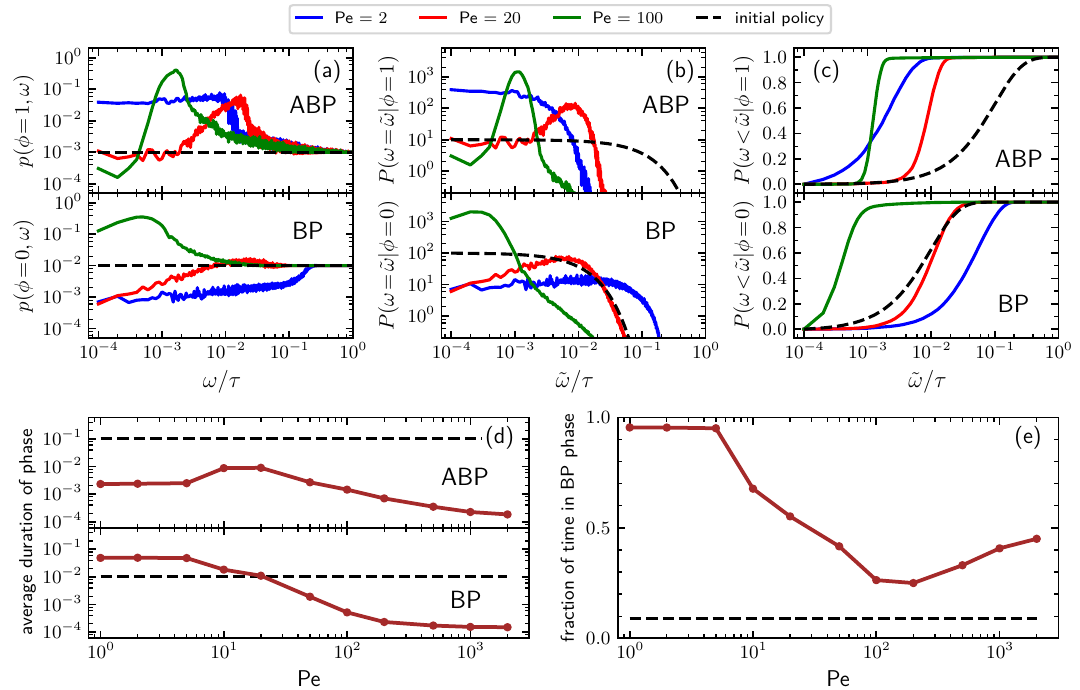}
\caption{
  \textbf{(a)} Probabilities of switching from BP to ABP motion (lower panel) and from ABP to BP motion (upper panel) as a function of the phase duration and for different P\'eclet numbers. Data are obtained after $10^3$ episodes and averaged over $N=5\cdot10^3$ independent particles;
  \textbf{(b)} Distribution of phase duration for different P\'eclet numbers (BP phase, lower panel -- ABP phase upper panel);
  \textbf{(c)} Cumulative distribution of the phase duration for different P\'eclet numbers;
  \textbf{(d)} Average duration of BP and ABP phases as a function of the activity;
  \textbf{(e)} Average fraction of time spent in the passive phase as a function of the activity.
  In all panels, the black dashed line represents the corresponding observable as obtained from the initial policy. 
    }
\label{fig:2}
\end{figure*}

An important question is how the target-search efficiency depends on the activity of the particle.
To address this issue, we investigate how the average time to reach the target during the $10^3$-th episode varies when changing the P\'eclet number.
This is reported in the inset of Figure~\ref{fig:1}, which shows that the learning particle has performances comparable to those of a passive particle as long as the P\'eclet number is smaller than about $\text{Pe}\approx10$ and then the average time to reach the target decreases with increasing activity until it reaches a plateau for $\text{Pe} \gtrsim 200$.
Such a phenomenology is consistent with the results already found in Ref.~\cite{Kaur2023} and is intuitively understood as follows:
Since the typical distance covered by pure diffusion grows with time as $t^{1/2}$ while the one due to the self-propulsion grows about as $t$, for small activities, diffusion dominates the relocation process during the short phases.
On the other hand, having long active phases is not favorable because of the particle's inability to find the target when in the ABP phase.
Thus, at low P\'eclet numbers, the learning particle tries to maximize the time spent in the passive phase, with the resulting performances equivalent to those of a purely passive particle.
In contrast, at large P\'eclet the self-propulsion velocity is enough to allow relocation at a distance larger than the target size even for very short active phases and the better performances of the learning particle are in accordance with the idea that having an intermittent-search strategy is more efficient than having a simple diffusive process.

\begin{figure*}[t]
\includegraphics[width=1.0\textwidth]{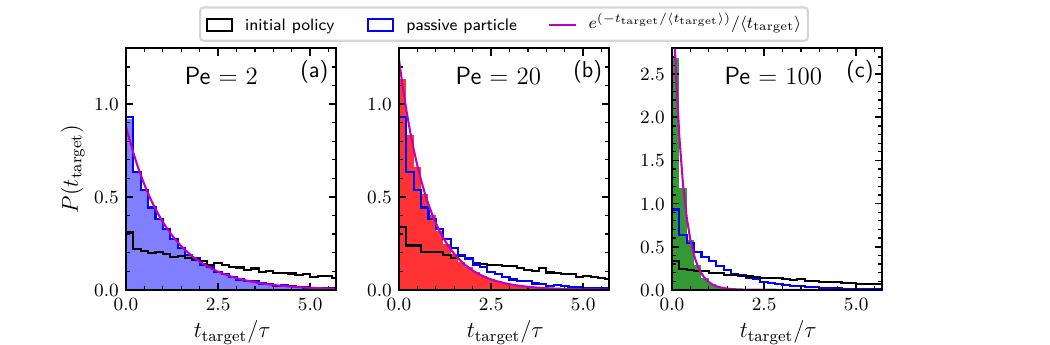}
\caption{
 Distribution of times needed to find the target collected during a time interval of length $10^6\tau$, for $\text{Pe}=2$, $20$, and $100$ (panels \textbf{a}, \textbf{b}, and \textbf{c} respectively). We consider here agents behaving according to the policies learned after $10^3$ episodes as reported in Figure~\ref{fig:2} (solid bars), to the initial policy prior to learning (black line), and completely passive particles (blue line). The magenta line is the exponential distribution having decay time given by the average time to find the target.
} \label{fig:3}
\end{figure*}

Additional insight into how the PS algorithm encodes learning successful strategies can be gained by directly investigating the policy, i.e. the probabilities of switching phase given the state.
This is done in Figure~\ref{fig:2} which reports these switching probabilities (panel a) and related observables as learned after $10^3$ episodes for $\text{Pe}=2$, $20$, and $100$, respectively corresponding to a low, intermediate, and high value of the activity.
Among the related observables, we report the probability of having a phase with a certain duration $\tilde{\omega}$ conditioned to being in phase $\phi$, $P(\omega=\tilde{\omega}|\phi)$ (Figure~\ref{fig:2}b) and the cumulative probability of having a phase duration shorter than $\tilde{\omega}$ conditioned to being in the phase $\phi$, $P(\omega<\tilde{\omega}|\phi)$ (Figure~\ref{fig:2}c).
These quantities can be obtained directly from the switching probabilities $p(s)=p(\phi,\omega)$ related to a given state $s=(\phi,\omega)$ as
\begin{equation}
P(\omega=\tilde{\omega}|\phi) = \dfrac{1}{\Delta t} \, \prod_{n'=1}^{\tilde{n}-1} \left[ 1-p(\phi,n' \Delta t) \right] \; p(\phi,\tilde{n}\Delta t) \; ,
\end{equation}
where we discretized the time introducing the integer variable $n = \omega/\Delta t$ and the factor $1/\Delta t$ in front of the right-hand side accounts for the correct normalization, and
\begin{equation}
P(\omega<\tilde{\omega}|\phi) = \sum_{n'=1}^{\tilde{n}-1} P(\omega' = n' \Delta t|\phi) \, \Delta t \; .
\end{equation}
Further observables reported in Figure~\ref{fig:2} are the average duration of a phase (panel d) and the fraction of time spent in the passive phase (panel e) as a function of the P\'eclet number.
However, before discussing the details of the learned policies, it is important to clarify that, for large enough $\omega$'s, the value of the switching probability $p(s)$ always drops to the corresponding value in the initial (arbitrary) policy.
In fact, the longer the phase duration $\omega$ of a given state $s=(\phi,\omega)$, the more rarely this state is visited during the learning process, with the frequency of these visits depending on the switching probabilities associated with the states $s=(\phi,\omega')$ having the same phase $\phi$ and lower phase duration $\omega'<\omega$.
This results in practical limitation in sampling states with a large phase duration.
In spite of this issue, the target-search abilities of the trained agent are not affected since the rarer it is to visit a given state, the smaller the contribution of the action following that state to the overall performances of the particle.

For low activity ($\text{Pe}=2$), the probability of switching from the passive to the active phase decreases from the value $10^{-2}$ corresponding to the initial policy to a value of about $10^{-3}$, slightly increasing with the duration of the ABP phase until it quickly converges to the initial policy value for a duration of the phase larger than about $10^{-1} \tau$ (see Figure~\ref{fig:2}a, lower panel).
On the other hand, the probability of switching from the active to the passive phase (Figure~\ref{fig:2}a, upper panel) increases from the initial policy's value $10^{-3}$ to about $4 \cdot 10^{-2}$ and drops to the initial policy for a duration of the phase larger than about $10^{-2} \tau$.
These results, together with the corresponding ones in panels b and c, indicate that, for low activity, the trained particle prefers to alternate relatively long passive phases with short active ones, confirming the previously mentioned expectations.
For $\text{Pe}=20$, the probability of having a phase with a certain duration $\omega$ of a given phase shows a peak at around $\omega = 10^{-2}\tau$ both when conditioned to be in the ABP phase ($\phi=1$) and in the BP one ($\phi=0$), see Figure~\ref{fig:2}b.
Consequently, for this value of the activity, the best strategy consists of alternating between active and passive phases both having a typical duration of about $10^{-2}\tau$ (see also Figure~\ref{fig:2}d), with the duration of the passive phase having a larger variance as indicated by the fact that the peak of the distribution conditioned to being in the active phase is narrower than the one of the distribution in the passive phase.
We stress that, because of its self-propulsion, an ABP with $\text{Pe}=20$ and $\ell^*=1$ in a time interval of $10^{-2}\tau$ covers a typical distance of about $0.2L$ which is twice the target diameter.
Finally, for $\text{Pe}=100$, Figure~\ref{fig:2} shows that the distribution of phase durations displays a sharp peak at around $10^{-3}\tau$ for the ABP phase and a rather broad peak at a few integration time steps.
Concomitantly, the learned strategy alternates between very short active phases with an average duration of about $1.4 \cdot 10^{-3}\tau$ and even shorter passive phases lasting about $0.5 \cdot 10^{-3}\tau$ on average.
In this case, the typical distance traveled during the active phase because of the particle's self-propulsion is about $0.14L$ which is of the same order as the one registered in the case of $\text{Pe}=20$ even though the activity is now $5$ times larger.

It is interesting to note that, as reported in Figure~\ref{fig:2}d, while the average passive phase duration monotonically decreases with the P\'eclet number, its counterpart for the active phase has a non-monotonic behavior that can be rationalized as follows:
Both at large and low values of the activity the ABP phases are very short but for two different reasons.
At low activity, these are short because active relocation to a distance greater than the target size would require too much time and the agent responds by minimizing the time spent in this phase.
In contrast, for large activity, very short active phases are already sufficient to allow the particle to relocate elsewhere in the simulation box and improve the target-search performances of the smart particle.
For intermediate P\'eclet numbers, the agent instead finds an optimal duration of the active phase reflecting the compromise between the utility of active phases for quick relocation and the fact that during these phases the target cannot be detected.
The effect of the monotonic and non-monotonic behaviors of the average duration of respectively the passive and the active phase, also results in a non-monotonic behavior of the fraction of time spent in the passive phase.
In fact, this quantity is close to $1$ for small P\'eclet numbers, decreases to about $0.25$ for $\text{Pe} \approx 100$, and then increases again to $0.5$ which is the value expected for extremely large levels of activity, see Figure~\ref{fig:2}e.

Finally, Figure~\ref{fig:3} shows the distributions of times needed to find the target by agents adopting the learned policies previously discussed for $\text{Pe}=2$, $20$, and $100$.
These distributions are exponential, meaning that the kinetics is completely characterized by the mean first-passage time.
For a comparison, the distributions obtained by a searching particle adopting the initial policy and by a completely passive particle are also reported.
Concerning the results obtained by adopting the initial policy, note that, even if the policy remains the same, in principle the resulting distribution depends also on the value of the activity.
However, this dependence appears to be very weak, as revealed by the similar behavior of the distribution corresponding to the three different P\'eclet numbers.
For low activity ($\text{Pe}=2$) the distribution of the searching times related to the learned policy is very similar to that of a simple BP, confirming the passive-like behavior of the agents in this P\'eclet regime.
As expected, increasing the activity, the distribution for the optimized particle becomes more and more narrow.
In particular, for $\text{Pe}=100$, the large majority of targets is found within the unit time $\tau$.

\section*{Conclusions}

In summary, we have introduced a RL-based method to probe the potential of smart microswimmers in target-search problems involving targets of unknown positions in a homogeneous environment.
Specifically, we applied a Projective Simulation based approach~\cite{munozgil2023} to micro-particles able to perform either passive or active Brownian motion and to switch from one to the other on the basis of a probabilistic policy.
Our findings demonstrate that, during repeated learning episodes, the agent optimizes its target-search performance and that the optimal policy strongly depends on the magnitude of the self-propulsion during the active phase.
For low activity, the behavior of the smart particle is similar to that of a completely passive particle while, for large activity, the agent takes advantage of the active phases to quickly cover more ground and increase the target-finding odds.
More in detail, the duration of the passive phases decreases with increasing P\'eclet number while the duration of the active phases displays a non-monotonic behavior with a maximum at intermediate P\'eclet numbers.
The proposed model is inspired by the intermittent search strategies developed by B{\'e}nichou and coworkers~\cite{Benichou2011,Benichou2005,Benichou2006}.
In this framework, an exponential distribution of phases is assumed to allow analytical tractability and in agreement with some experimental observations~\cite{Benichou2011}.
However, in our case, the duration of a phase is part of the state sensed by the agent, meaning that the agent is endowed with some sort of temporal memory.
Consequently, distributions of phases different from the exponential one may arise, which is indeed what is observed in the learned policies.
Our results complement and extend those of a previous study based on a genetic algorithm~\cite{Kaur2023} and demonstrate that also reinforcement learning is a powerful tool to investigate target-search problems for agents undergoing a stochastic dynamics.

With respect to previous literature on stochastic target search~\cite{Viswanathan2011,Viswanathan1999,Viswanathan2008,Benichou2011,Benichou2005,Benichou2006,Loverdo2009,munozgil2023}, which mainly applies to generic scenarios, our investigation is more focused on the microscopic world, namely we are interested in natural or artificial microswimmers.
This is the main reason to resort to the active Brownian particle model.
In fact, this model, besides being the paradigmatic model in the framework of non-equilibrium dynamics~\cite{Cates2012,Fodor2016,Fodor2018,Caraglio2022}, also provides a faithful representation of the behavior of artificial microswimmers such as the Janus particles~\citep{Bechinger2016}.
Remarkably, nowadays it is already possible to perform experiments in which the activity of artificial microswimmers is controlled by an external illuminating system~\cite{Muinos-Landin2021}.
Thus, the target-search strategies developed in the present manuscript can potentially be tested in a laboratory.
Furthermore, the \textit{intermittent} active Brownian dynamics that we introduce in the Model section can be also considered, in the case of relatively large activity and persistence, as a first proxy for the run-and-tumble dynamics which is the typical theoretical model describing the motion of bacteria~\cite{Berg2004,Bechinger2016,Santra2020}.

The proposed framework offers new insight into target-search problems in homogeneous enviroments and paves the way to further research.
In particular, it can be leveraged to explore more complex scenarios such as, for instance, target search with resetting events~\cite{Evans2011,Kusmierz2014,Kumar2020}, multiple and/or motile targets problems~\cite{Benichou2011}, or searchers with multiple migration modes, the latter being relevant to dendritic cells searching for infections~\cite{Song2023}.
Moreover, other possible developments, particularly relevant for the envisioned medical and environmental application of smart active particles, entail heterogeneous environments involving the presence of obstacles, boundaries, and energy barriers~\cite{Volpe2017,Zanovello2021,Zanovello2021b,Zanovello2023}.
Finally, endowing the agent with a limited memory of the recently visited locations~\cite{Meyer2021} or with the ability to sense directional cues coming from the target itself, may also be an extension going in the direction of better modelling biological microswimmers.

\section*{Methods}

To identify effective target-search strategies, we used the RL algorithm Projective Simulation (PS), which was originally created as a platform for the design of autonomous quantum learning agents~\cite{Briegel2012} and was shown to have competitive performance also in classical RL problems~\cite{Mautner2015,Boyajian2020}.

The core idea of this algorithm is to use the notion of a particular kind of memory, called \textit{episodic and compositional memory} (ECM) which is mathematically described by a graph connecting units called \textit{clips}.
Clips can be either percept or decision units, corresponding to states and actions respectively, or a combination of those.
We design our target-search problem as a Markov decision process~\cite{Sutton2018}, i.e. at each learning step, the agent is in some state $s$, takes an action $a$ according to a policy defined by the conditional probabilities $\pi(a | s)$, and receives a reward $\mathcal{R}$ as a consequence of this action.
In such a case, the ECM structure consists of a layer of states fully connected with a layer of actions.
Each edge of the graph, i.e each state-action pair $(s,a)$, is assigned with a real-value weight $h(s,a)$, called the $h-$value, which determines the policy according to
\begin{equation}
\pi(a|s) = \dfrac{h(s,a)}{\sum_{a' \in \mathcal{A}} h(s,a')} \; , 
\end{equation}
where $\mathcal{A}$ represents the set of all possible actions.
Furthermore, a non-negative glow value $g(s,a)$ stores the information on which and, implicitly, how frequently state-action pairs have been visited during the learning process.
Such information is then exploited when updating the policy with the goal of maximizing the total expected reward.

This last feature of the PS algorithm makes it particularly apt to solve our target-search problem:
Indeed, on average, the equations of motion~(\ref{eom1}-\ref{eom3}) have to be iterated a large number of times before a target is found and the agent obtains its reward.
Consequently, the reward signal is very sparse and has only a very low correlation with the particular state-action pair encountered when the target is found.
Approaches taking into account long sequences of visited state-action pairs, as the PS algorithm, should then be preferred with respect to typical action-value methods such as one-step Q-learning or SARSA~\cite{Sutton2018}.

Applying the PS framework to the model illustrated in the dedicated section and taking into consideration that in our case the action $a$ can be described as a binary variable, with $a=1$ corresponding to a switch of the phase (passive or directed motion) and $a=0$ to maintaining the current phase, a single learning step consists of the following operation:
\begin{itemize}
\item Given the current state $s_t = (\phi_t,\omega_t)$, the probability of switching phase $p_t$ is determined as $$p_t = \pi(a_t = 1|s_t) = h(s_t,1)/[h(s_t,0) + h(s_t,1)]$$ and the next phase $\phi_{t+1}$ is selected accordingly;
\item The glow matrix is damped following the update rule $G \leftarrow (1-\eta)G$, where $\eta$ is called the glow parameter and determines how much a delayed reward should be discounted;
\item The glow matrix is updated by adding a unit to the visited state-action pair, $g(s_t,a_t) \leftarrow g(s_t,a_t)+1$;
\item The position and the direction of the particle evolve according to Eqs.~(\ref{eom2}-\ref{eom3});
\item The matrix of $h$-values is updated according to the learning rule of the PS model, $H \leftarrow (1-\gamma) H + \gamma H_0 + \mathcal{R} \, G$, where $\mathcal{R}$ is the reward being zero if no target is found by the particle located at the updated position and $1$ otherwise.
Here, $\gamma$ is called the damping parameter and specifies how quickly the $H$ matrix returns to an initial matrix $H_0$.
\end{itemize}
The initial policy is such that the probabilities of switching phase are $10^{-2}$ and $10^{-3}$ when being in the passive and in the active phase respectively.
This is obtained by setting, for each $t$,
$h_0(s_t,a_t=1)=10^{-2}$ and $h_0(s_t,a_t=0)=(1-10^{-2})$ if the state is in a passive phase, and $h_0(s_t,a_t=1)=10^{-3}$ and $h_0(s_t,a_t=0)=(1-10^{-3})$ if the state is in an active phase.
All the terms of the $G$ matrix are initialized to zero at the beginning of each episode.

We set the integration time step to $\Delta t = 10^{-4}\tau$ and, to have a finite set of states, we limit the duration of a given phase $\omega$ to be not longer than $\tau$.
This results in a total of $2 \cdot 10^4$ states $s_t=(\phi_t, \omega_t)$, being $\phi_t = 0,1$ (see Model section) and $\omega_t = 1,\ldots,10^4$.
The glow and the damping parameters are considered hyperparameters of the model and, for each value of the activity $\text{Pe}$ and of the persistence $\ell^*$, are adjusted to obtain the best learning performances.
Their values are reported in Table~\ref{tab:1}.
Finally, to investigate how the learning process evolves, we split the whole process into several episodes, each lasting $20\tau$.
At the beginning of each episode, each element of the glow matrix is initialized to zero.

\begin{table}
\begin{center}
\begin{tabular}{|c|ccccc|}
\hline 
$\text{Pe}$ & $\leq5$    & $10$      & $20$      & $50$      & $\geq100$  \\ 
\hline 
$\gamma$    & $10^{-7}$  & $10^{-6}$ & $10^{-6}$ & $10^{-6}$ & $10^{-5}$ \\  
$\eta$      & $10^{-2}$  & $10^{-3}$ & $10^{-3}$ & $10^{-2}$ & $10^{-2}$  \\ 
\hline 
\end{tabular}
\end{center}
\caption{Hyperparameters used to obtain the results presented in the present work. \label{tab:1}}
\end{table}

\section*{Acknowledgements}

H.K. acknowledges funding from the European Union's Horizon 2020 research and innovation programme under the Marie Sk\l{}odowska-Curie grant agreement No 847476;
M.C. is supported by FWF: P 35872-N;
T.F. acknowledges funding by FWF: P 35580-N;
A.L. and H.J.B. acknowledge support by the Volkswagen Foundation (Az:97721);
H.J.B. acknowledges funding from  FWF through SFB BeyondC F7102, and the European Research Council (ERC, Quant AI, Project No. 10105529).
G.M-G also acknowledges funding from the European Union.
\newpage
Views and opinions expressed are however those of the author(s) only and do not necessarily reflect those of the European Union, the European Research Council or the European Research Executive Agency . Neither the European Union nor the granting authorities can be held responsible for them.



%

\end{document}